\documentclass[twocolumn,superscriptaddress,showpacs,prl]{revtex4}

\usepackage{graphicx}
\usepackage{dcolumn}
\usepackage{bm}
\usepackage{color}

\begin{document}

\preprint{APS/123-QED}

\title{Doping Dependent Changes in Nitrogen 2$p$ States in the Diluted Magnetic Semiconductor Ga$_{1-x}$Cr$_{x}$N}
\author{T. Takeuchi}
\affiliation{Department of Applied Physics, Tokyo University of Science, Tokyo 162-8581, Japan.}
\affiliation{RIKEN/SPring-8, Kouto 1-1-1, Mikazuki-cho, Sayo-gun, Hyogo 679-5148, Japan.}
\author{Y. Harada}
\affiliation{RIKEN/SPring-8, Kouto 1-1-1, Mikazuki-cho, Sayo-gun, Hyogo 679-5148, Japan.}
\author{T. Tokushima}
\affiliation{RIKEN/SPring-8, Kouto 1-1-1, Mikazuki-cho, Sayo-gun, Hyogo 679-5148, Japan.}
\author{M. Taguchi}
\affiliation{RIKEN/SPring-8, Kouto 1-1-1, Mikazuki-cho, Sayo-gun, Hyogo 679-5148, Japan.}
\author{Y. Takata}
\affiliation{RIKEN/SPring-8, Kouto 1-1-1, Mikazuki-cho, Sayo-gun, Hyogo 679-5148, Japan.}
\author{A. Chainani}
\affiliation{RIKEN/SPring-8, Kouto 1-1-1, Mikazuki-cho, Sayo-gun, Hyogo 679-5148, Japan.}
\author{J. J. Kim}
\affiliation{Center for Interdisciplinary Research, Tohoku University, Sendai 980-8578, Japan.}
\author{H. Makino}
\affiliation{Institute for Material Research, Tohoku University, Sendai 980-8577, Japan.}
\author{T. Yao}
\affiliation{Center for Interdisciplinary Research, Tohoku University, Sendai 980-8578, Japan.}
\affiliation{Institute for Material Research, Tohoku University, Sendai 980-8577, Japan.}
\author{T. Yamamoto}
\affiliation{Department of Electronic and Photonic Systems Engineering, Kochi University of Technology, Kochi 782-8502, Japan.}
\author{T. Tsukamoto}
\affiliation{Department of Applied Physics, Tokyo University of Science, Tokyo 162-8581, Japan.}
\author{S. Shin}
\affiliation{RIKEN/SPring-8, Kouto 1-1-1, Mikazuki-cho, Sayo-gun, Hyogo 679-5148, Japan.}
\affiliation{Institute for Solid State Physics, The University of Tokyo, Chiba 277-8581, Japan.}
\author{K. Kobayashi}
\affiliation{JASRI/SPring-8, Kouto 1-1-1, Mikazuki-cho, Sayo-gun, Hyogo 679-5198, Japan.}
\date{\today}

\begin{abstract}
We study the electronic structure of the recently discovered diluted magnetic semiconductor Ga$_{1-x}$Cr$_{x}$N ($x$ = 0.01-0.10). A systematic study of the changes in the $occupied$ and $unoccupied$ ligand (N) partial density of states (DOS) of the host lattice is carried out using N 1$s$ soft x-ray emission and absorption spectroscopy, respectively.  X-ray absorption measurements confirm the wurtzite N 2$p$ DOS and substitutional doping of Cr into Ga-sites. Coupled changes in the $occupied$ and $unoccupied$ N 2$p$ character DOS of Ga$_{1-x}$Cr$_{x}$N identify states responsible for ferromagnetism consistent with band structure calculations.
\end{abstract}

\pacs{75.50.Pp, 71.20.-b, 78.70.En, 78.70.Dm}
\maketitle
To achieve practical spintronics,\cite{Wolf_Science 294} it is important to realize diluted magnetic semiconductors (DMS) with a ferromagnetic Curie temperature, $T_C$, higher than room temperature.  Beginning with the work of hole doping induced ferromagnetism in In$_{1-x}$Mn$_{x}$As\cite{Munekata_PRL63} and Ga$_{1-x}$Mn$_{x}$As\cite{Ohno_APL69} systems, there has been extensive effort to synthesize higher $T_C$ materials, particularly using molecular beam epitaxy (MBE) based methods. Enormous progress has been achieved in the last decade in creating materials which exhibit fascinating properties such as 'spin memory'\cite{Kikkawa_Science277}, 'photo-induced'\cite{Koshihara_PRL78} and 'electric-field induced'\cite{Ohno_Nature408} magnetism, in addition to the doping-induced ferromagnetism. Theoretically, Dietl {\it et al}. showed the importance of Zener's $p$-$d$ exchange interaction for the carrier mediated ferromagnetic order in DMS materials\cite{Dietl_Science287}. They also predicted that the ferromagnetic $T_C$ can be enhanced to above room temperature with Mn doping in GaN and ZnO, fuelling the prospects of carrier mediated ferromagnetism in semiconductors\cite{Garf_PSSB239}. Combining first principles band structure calculations in the local spin-density approximation with a Heisenberg model to describe magnetic properties, Sato {\it et al.} predicted that Cr substitution instead of Mn should lead to higher ferromagnetic $T_C$'s than the Mn doped III-V systems.\cite{Sato_EPL61} Experimentally, significant effort is being devoted to synthesize room temperature ferromagnets based on the II-VI and III-V semiconductors with transition metal substitutions.\cite{Ueda_APL79,Park_APL,Yamada_JAP91,Hashimoto_JCG251,Suemasu_PSSC0,Kim_PSS0,Liu_condmat} While Co substitution in ZnO has met with limited success\cite{Ueda_APL79}, recent reports of Cr substitution in GaAs and GaN have demonstrated ferromagnetism\cite{Park_APL,Yamada_JAP91,Hashimoto_JCG251,Suemasu_PSSC0,Kim_PSS0,Liu_condmat}. Single crystal Cr-doped GaN with a $T_C$ = 280 K\cite{Park_APL} was soon followed by MBE growth of Ga$_{1-x}$Cr$_{x}$N with a $T_C$ in excess of 350 K\cite{Hashimoto_JCG251,Suemasu_PSSC0,Kim_PSS0,Liu_condmat}.

In establishing ferromagnetism, it is important to investigate segregation and/or clustering of impurity phases at the nanoscale as suggested by theoretical studies\cite{Schilfgaarde_PRB63}. This was recently shown from experimental studies of the Co doped TiO$_2$ system\cite{Kim_APL81}. Alternatively, the role of a combination of substitutional and interstitial doping could also be important, as in Ga$_{1-x}$Mn$_x$As, in which the annealing dependence of Ga$_{1-x}$Mn$_x$As can be explained on the basis of migration of Mn interstitials to substitutional sites\cite{Mahadevan_PRB68}. In another scenario, the annealing can cause evaporation of excess As, leading to a transformation of paramagnetic Mn ions to ferromagnetic Mn species\cite{Ishiwata_PRB65}. However, the enhancement of ferromagnetic $T_C$'s is always accompanied by an increase in charge carrier density. This indicates modifications in the $occupied$ and/or $unoccupied$ density of states which is suitably addressed by valence and conduction band spectroscopy. In this context, it is important to use spectroscopy to measure the changes in the density of states, and simultaneously ensuring that, the spectroscopy measures the intrinsic modifications and not spurious features due to impurities at the nanoscale.

In this work, we investigate the newly synthesized material Ga$_{1-x}$Cr$_{x}$N which shows room temperature ferromagnetism. We use soft x-ray emission spectroscopy (SXES) and x-ray absorption spectroscopy (XAS) across the N 1$s$ threshold to probe doping dependent changes in the host lattice. SXES and XAS are complimentary techniques which probe site-selective angular momentum projected valence and conduction band DOS around an ion, respectively\cite{Nordgren_PST31,Stagarescu_PRB54,Smith_JVSTB16,Kotani_RMP73}. Another important reason to choose the N 1$s$ threshold for the present study is that GaN has significantly large N 2$p$ character DOS while the Ga 3$d$ DOS as probed by Ga $L_{\alpha}$ SXES is very low at and near the valence band maximum, as is well-known from earlier studies\cite{Stagarescu_PRB54,Smith_JVSTB16}. Our study provides direct evidence for Cr doping induced changes in the N 2$p$ character DOS of the $occupied$ and $unoccupied$ DOS in Ga$_{1-x}$Cr$_{x}$N .The present study also shows that the ligand derived $occupied$ DOS exhibit small but systematic changes coupled to changes in $unoccupied$ DOS.  The angular dependence of the XAS data\cite{Katsikini_APL69} is used to show that the N 2$p$ partial DOS is due to the wurtzite structure for $x$ = 0.0 to 0.10, except for the additional new doped states which are formed in the band gap of GaN. Cr 2$p$ XAS with cluster model calculations confirm the tetrahedral coordination of Cr in the Ga-site.

GaN and Cr-doped GaN films were grown by NH$_3$-MBE on ZnO substrates.  The details of growth and magnetic properties of Ga$_{1-x}$Cr$_{x}$N have been reported recently\cite{Kim_PSS0}.  SXES and XAS measurements were performed at the undulator beamline BL27SU in SPring-8 using linearly polarized light\cite{Ohashi_NIMA467}.  SXES spectra were recorded using a flat field type spectrometer equipped with CCD (Charge-Coupled device) detector.\cite{Tokushima_SRL9} XAS spectra were obtained by measuring the sample drain current. All the experiments reported were carried out at room temperature.  The energy resolution was set to 40 meV at 400 eV for XAS measurements, while the SXES resolution was 400 meV at 400 eV as determined from specular reflection of GaAs substrates.  Angle-dependent spectra were recorded with an accuracy of +/- 1 degree.

\begin{figure}
\includegraphics[width=.9\linewidth]{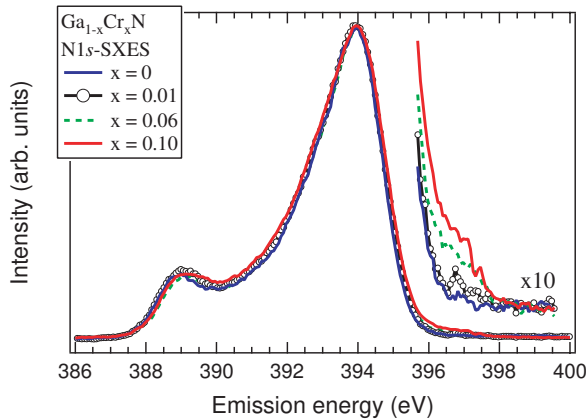}
\caption{(Color online) Doping dependence of N 1$s$ SXES spectra for Ga$_{1-x}$Cr$_{x}$N ($x$ = 0.0, 0.01, 0.06 and 0.10) using 401.4 eV photon energy, together with enlarged spectra around 397 eV emission energy.}
\end{figure}

\begin{figure}
\includegraphics[width=.9\linewidth]{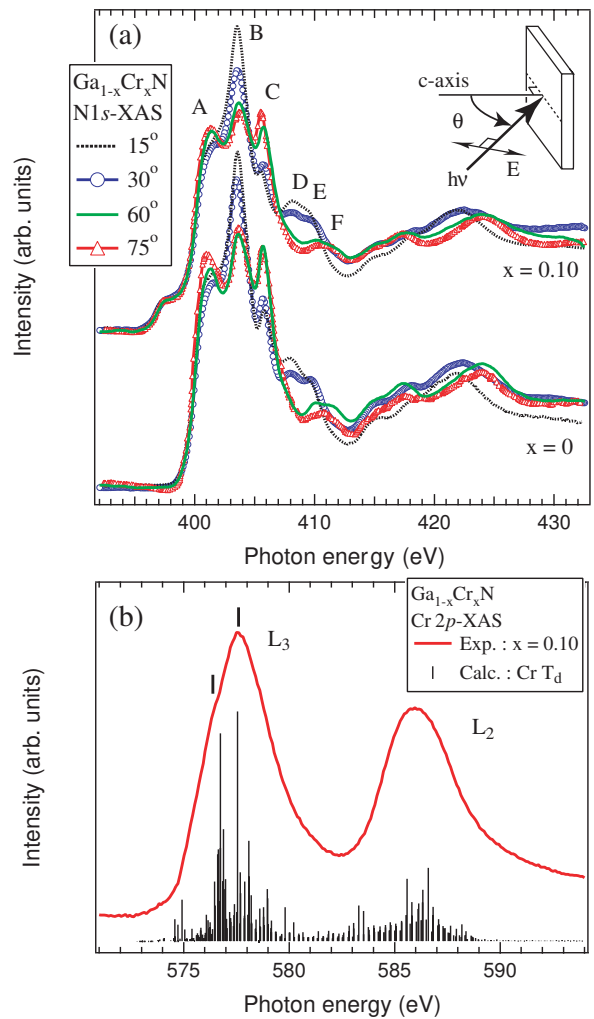}
\caption{(Color online) (a) Photon incidence angle dependence of N 1$s$ XAS spectra for (upper panel) Ga$_{1-x}$Cr$_{x}$N ($x$ = 0.10) and (lower panel) GaN. The upper inset defines the photon incidence angle. (b) Cr 2$p$ XAS spectrum for Ga$_{1-x}$Cr$_{x}$N ($x$ = 0.10) with cluster model calculations (line diagram) confirming tetrahedral coordination.}

\end{figure}
Figure 1 shows SXES spectra of Ga$_{1-x}$Cr$_{x}$N ($x$ = 0.0 - 0.10) excited with a photon energy of 401.4 eV, just across the N 1$s$ threshold to enhance the elemental specificity\cite{Nordgren_PST31,Stagarescu_PRB54,Smith_JVSTB16}. The spectra represent the N 2$p$ partial DOS in the valence band. The spectrum of pure GaN without doping exhibits two features, a high intensity feature around 394 eV and a weak feature around 389 eV. The spectrum is very similar to that reported by Stagarescu {\it et al.} at an incident photon energy of $\sim$400 eV\cite{Stagarescu_PRB54}. For the Cr-doped spectra, the overall spectral shape remains very similar, except for a low intensity feature that grows systematically just above the valence band maximum of GaN. To see this clearly, the spectra are plotted on an expanded scale in the inset. We clearly see that the Cr doping induces an additional feature around 397 eV emission energy, which systematically grows with Cr content. It represents the N 2$p$ states hybridized with the Cr 3$d$ states in the occupied valence band of Ga$_{1-x}$Cr$_{x}$N. In contrast, the N 2$p$ partial DOS shows a shift of the leading edge with respect to the valence band maximum for Al substitution in Al$_x$Ga$_{1-x}$N\cite{Smith_JVSTB16}.

Having studied the occupied N 2$p$ partial DOS, we measured N 1$s$ to 2$p$ XAS spectra of Ga$_{1-x}$Cr$_{x}$N ($x$ = 0.0 - 0.10) to probe the conduction band DOS. The XAS spectra were obtained with linearly polarized radiation, corresponding to the geometry shown in the inset to Fig. 2(a).  Figure 2(a) shows the incident angle dependence of the XAS spectra for the highest Cr content ($x$ = 0.10) and pure GaN compositions.  The angular dependence for the intermediate compositions has also been measured to be similar to Fig 2(a).  All the spectra are normalized for area under the curve from $h\nu$ = 395 eV to 418 eV. The angular dependence represents the orbital character of the spectral features in GaN. In Fig 2(a), the sharp features labeled A and C at 401 and $\sim$406 eV, and a broad weak feature F, centered at about 411 eV are due to N 2$p_z$ states. The features labeled B, D and E at 403.5, 408 and 410~eV, respectively are due to N 2$p_x$ and 2$p_y$ states of the wurtzite structure. The observed angular dependence of the N 1$s$ to 2$p$ XAS spectrum for GaN is well-established as a fingerprint of the N 2$p$ character DOS of the wurtzite structure\cite{Katsikini_APL69}, ruling out cubic CrN. The fact that the angle-dependent spectra are nearly identical for GaN and with up to 10\% Cr doping (Fig. 2(a)), feature for feature, indicates that the site-selective angular momentum projected N 2$p$ DOS is not modified with Cr doping, except for the new states formed in the band gap of GaN, just below the conduction band minimum. This confirms absence of impurity nitride phase. To rule out Cr clustering or dimer formation, we have carried out Cr 2$p$ to 3$d$ XAS, and the spectrum for the highest doping concentration of 10\% Cr is in Fig. 2(b). The experimental spectrum is consistent with a cluster calculation for trivalent Cr in the tetrahedral geometry of the wurtzite structure using three configurations: 3$d^3$, 3$d^4\underline{L}$ and 3$d^5\underline{L}^2$, where $\underline{L}$ is hole in ligand (N) states \cite{Taguchi_PRB63}. The ground state consists of 26.4\% 3$d^3$, 54.1\% 3$d^4\underline{L}$ and 19.5\% 3$d^5\underline{L}^2$, indicating strong hybridization. The crystal field split features (tick marks in Fig. 2(b)) are discernible in the $L_3$ edge, as understood from the calculations. We have also tried trivalent Cr in octahedral geometry, but the calculations do not match the data as well. This confirms tetrahedral coordination of Cr ions in the host lattice, and hence indicates absence of Cr clustering or dimer formation.

\begin{figure}
\includegraphics[width=.9\linewidth]{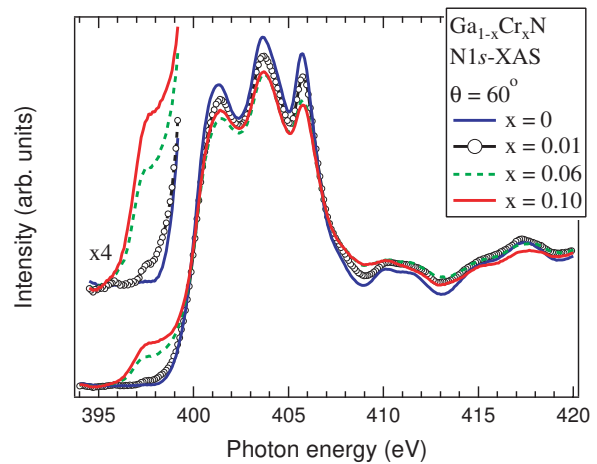}
\caption{(Color online) Doping dependence of N 1$s$ XAS spectra for Ga$_{1-x}$Cr$_{x}$N ($x$ = 0.0, 0.01, 0.06 and 0.10) together with enlarged spectra around 397 eV photon energy.  The spectra are normalized from 395 eV to 418 eV for area under the curve. All spectra are for 60$^\circ$ photon incidence angle.}
\end{figure}

The doping dependence ($x$ = 0.0 - 0.10) of the spectra for a particular angle (60$^\circ$) is shown in Fig. 3.  The spectra reveal a systematic growth of unoccupied states with Cr substitution, within the band gap of GaN. In order to elucidate the nature of these states, we compare the experimental occupied and unoccupied partial DOS with the calculated partial DOS obtained from results of our electronic band structure calculations for pure GaN and GaN doped with Cr (Ga$_{1-x}$Cr$_{x}$N), based on density functional theory (DFT) \cite{Kohn_PR140} with generalized gradient approximation (GGA) \cite{Perdew_ESS91}.  The cell is optimized with VASP \cite{Kresse_PRB54} based on a projector-augmented-wave (PAW) approach \cite{Kresse_PRB59} with the following parameters: $k$-spacing of 0.3/\AA, SCF convergence of 1.0 $\times$ 10$^{-5}$ eV, gradient convergence of 0.03 eV/\AA, and Gaussian smearing.  We study the electronic structures of GaN and Ga$_{1-x}$Cr$_{x}$N with periodic boundary conditions by generating supercells that contain the object of interest.  For a GaN:Cr supercell to represent Ga$_{0.9375}$Cr$_{0.0625}$N, we replace one of the 16 sites of Ga atoms by a Cr site in the supercell.

\begin{figure}
\includegraphics[width=.9\linewidth]{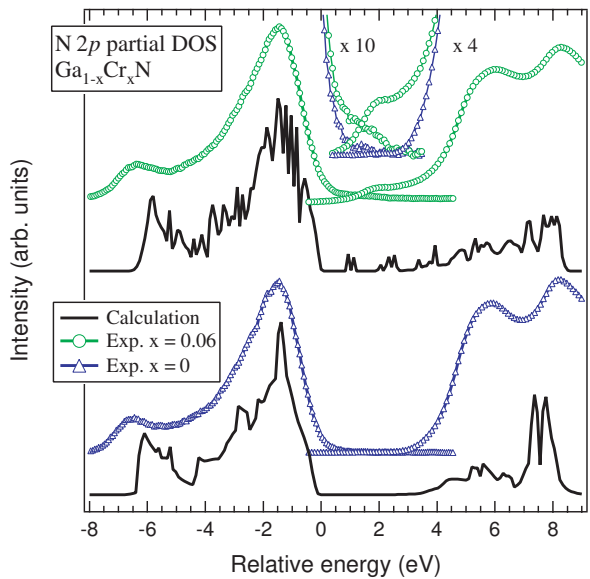}
\caption{(Color online) The N 2$p$ PDOS as calculated for Ga$_{1-x}$Cr$_{x}$N ($x$ = 0.0625) compared with Ga$_{1-x}$Cr$_{x}$N ($x$ = 0.06) N 1$s$ XAS and SXES spectra (upper panel), and similarly for GaN (lower panel). Inset shows changes in the band gap on Cr substitution.}
\end{figure}

Figure 4 shows the experimental data of GaN and Ga$_{0.94}$Cr$_{0.06}$N compared with the calculated N 2$p$ partial DOS. We adopt the same procedure as Stagarescu {\it et al.} to calibrate the energy scale of GaN and the doped materials, as compared to the photon energy scale\cite{Stagarescu_PRB54}. The overall results of the DFT calculations are consistent with the SXES and XAS spectra, although there are some remaining discrepancies for large positive relative energies (Fig.4).   The peak structure obtained in the band gap of GaN corresponds to the $e_g$ and $t_{2g}$ split up-spin ($e_\uparrow$ and $t_\uparrow$) and down spin ($e_\downarrow$ and $t_\downarrow$) states hybridized with the N 2$p$ states. Assuming a formal Cr$^{3+}$ state, the Fermi level is expected to lie in the up-spin $t_\uparrow$ band with one electron occupied to give a metallic ground state.  The real system is actually semiconducting, in the present case as well other reports\cite{Hashimoto_JCG251,Suemasu_PSSC0,Kim_PSS0,Liu_condmat}, but displays variable range hopping conductivity as reported recently\cite{Liu_condmat}.  The observed changes in SXES correspond to fully $occupied$ $e_\uparrow$ and the partially $occupied$ $t_\uparrow$ states. The lowest lying $unoccupied$ states in XAS would thus be expected to have $t_\uparrow$-spin band character consisting of $d_{xy}$, $d_{yz}$ and $d_{zx}$ character, with higher lying $e_\downarrow$ and $t_\downarrow$. The lac$k$ of angular dependence in the lowest lying $unoccupied$ states formed due to Cr-substitution confirms this picture. A recent self-consistent linear muffin tin orbital calculation of Cr doped GaN, shows a similar peak structure in the gap for a ferromagnetic ground state\cite{Das_condmat}.

In a recent x-ray magnetic circular dichroism experiment on Ga$_{1-x}$Mn$_x$As\cite{Keavey_PRL91}, it was shown that the As 4$s$ signal exhibits a magnetic moment antiparallel to the Mn moment while the Ga 4$s$ moment is parallel to the Mn moment. This experiment provides conclusive evidence of carrier mediated ferromagnetism involving the host As-lattice. It would be important to study magnetic circular dichroism, and also spin-polarized photoemission, of Ga$_{1-x}$Cr$_{x}$N to determine the spin polarization of the doping dependent changes in the $occupied$ and $unoccupied$ DOS.

In conclusion, we have performed N 1$s$ SXES and XAS and Cr 2$p$ XAS to study the electronic structure of DMS Ga$_{1-x}$Cr$_{x}$N ($x$ = 0.01 - 0.10) and GaN. The study confirms the wurtzite N 2$p$ DOS and on-site substitution of Cr for Ga from XAS. The valence and conduction band show systematic changes with the new doped states of N 2$p$ character hybridized with Cr 3$d$ states, placed in the band gap of GaN.  These doped states are responsible for the ferromagnetism in Ga$_{1-x}$Cr$_{x}$N.

We thank Dr. H. Ohashi and Dr. Y. Tamenori for valuable technical assistance at BL27SU in SPring-8.  This experiment was carried out in SPring-8 with the approval of the Japan Synchrotron Radiation Research Institute (JASRI) (Proposal Nos. 2003A0648-NS1-np and 2003B0760-NSa-np).


\end{document}